\documentclass[12pt]{iopart}
\usepackage{iopams}
\usepackage{graphicx}
\newcommand{\hb}[1]{\hat{\bf #1}}
\newcommand{\B}[1]{{\bf #1}}
\begin{document}

\title{Is the number of Photons  a Classical Invariant?}

\author{J.~E.~Avron, E.~Berg, D.~Goldsmith and A.~Gordon}
\address{ Department of Physics, Technion, 32000 Haifa, Israel}

\begin{abstract}
We describe an apparent puzzle in classical electrodynamics and
its resolution. It is concerned with the  Lorentz invariance of
the classical analog of the number of photons.
\end{abstract}

\section{Introduction}

Photon are quantum objects and {\it a priori} have no business in
classical electrodynamics. So, what can one possibly mean by the
question: 'Is the number of Photons  a Classical Invariant?'

Consider a box filled with monochromatic radiation
of frequency $\omega$.  If $U$ denotes the total electromagnetic
energy in the box, then, the right hand side of
\begin{equation}
\label{N}
\hbar N=\frac{U}{\omega}
\end{equation}
is a purely classical quantity. The left hand side gives the
 interpretation and quantization of this quantity, namely, that
it counts the number of photons, $N$, in units of
${\hbar}$. What is the classical significance of $U/\omega$?

In quantum mechanics the number of photons is quantized. As such, it must
be Lorentz invariant, for under Lorentz transformations that are close to
the identity, it can only change by a little, and since it is quantized it
cannot change at all. This implies that the number of photons must be a
Lorentz invariant, even under Lorentz transformations that are far from
the identity. With this hindsight, and since Lorentz invariance is a
classical concept, one learns that the classical significance of $U/\omega$
is its Lorentz invariance.

Since neither the energy nor the frequency are Lorentz invariants,
the Lorentz invariance of the ratio is not manifest, and as we
shall see is actually a rather subtle. If, indeed, the Lorentz
invariance of the ratio holds by a direct classical argument,
without recourse to the quantization argument above, one can
rediscover, and to some extent also motivate the existence of
photons on purely classical grounds. This approach has its
limitations, of course. One still needs quantum mechanics to
understand quantization, and $\hbar$  to actually count photons.

After the preprint of this paper was posted on the Los Alamos
electronic archive,  Prof. Andrew Zangwill, drew our attention to
a paper by  Ya. B. Zeldovich  in \cite{Zeldovich} who considered
this problem  in  1966.  In section 7 of this paper, we shall
discuss Zeldovich paper in some detail.

\section{The Puzzle}

Here is, what appears to be a reasonable
 calculation of how equation~(\ref{N}) Lorentz transforms.
 In this calculation   $\frac{U}{\omega}$
turns out not to be Lorentz invariant.

Consider a linearly polarized, plane monochromatic wave of frequency $\omega$
traveling in the $\hat {\bf x}$ direction.
The  electric and magnetic  fields are:
\begin{equation}
{\bf E}=E_0\,\cos(kx-\omega t)\, \hat {\bf y},\quad \bf B\rm=E_0\,\cos(kx-\omega
t)\, \hat {\bf z}.
\end{equation}
The electromagnetic energy density  is
\begin{equation}
\frac{1}{8\pi}\left({\bf E}^{2}+{\bf B}^{2}\right)= \frac{E_0^
2}{4\pi} \,\cos^{2}(kx-\omega t).
\end{equation}
Consider  a {\em fictitious} rectangular box of proper length $L$,
aligned with the $x$ axis,
whose cross
section  is $A$. Suppose that the length of the
box is much larger than the wave length of the radiation. The
total energy in the box is
\begin{equation}
\label{eq:energy}
U=\frac{A E_0^2}{4\pi}\int_0^L\, dx\,\cos^{2}(kx-\omega t)\approx
 \frac{A\,L\, E_0^2}{8\pi}.
 \end{equation}
The number of photons in this box, according to equation~(\ref{N}), appears to be
\begin{equation}
\label{eq:n1}
 \hbar N = \frac{U}{\omega}= \frac{E_0^ 2 AL}{8\pi\omega}.
\end{equation}
Now, let us compute the number of photons, $N'$, in the same
box, but as viewed
in a frame, $S'$, moving with
velocity $v$ along the $x$ axis. In $S'$, the electric field amplitude
\cite{landau} is:
\begin{equation}
\label{eq:new_e}
E_0' = \frac{E_y-(v/c)B_z}{\sqrt{1-(v/c)^2}}=\frac{E_0-(v/c)E_0}
{\sqrt{1-(v/c)^2}}=E_0\sqrt{\frac{1-v/c}{1+v/c}}
\end{equation}
 The length of the box experiences
Lorentz contraction and is now:
\begin{equation}
\label{eq:new_L}
 L' = L\sqrt{1-(v/c)^2}
\end{equation}
The electromagnetic energy in the box in the moving frame is therefore
\begin{equation}
\label{eq:new_energy}
 U' \approx \frac{(E_0') ^ 2}{8\pi}AL'=
 \frac{E_0^ 2}{8\pi}\frac{1-v/c}{1+v/c}AL\sqrt{1-(v/c)^2}.
\end{equation}
Now, $ \omega $ is transformed according to the Doppler formula
 \cite{berkeley}:
\begin{equation}
\label{eq:new_omega}
\omega'=\omega\sqrt{\frac{1-v/c}{1+v/c}}
\end{equation}
Hence  the number of photons in the moving  box appears to be:
\begin{equation}
\label{eq:new_n1}
 \hbar N' = \frac{U'}{\omega'}\approx \frac{E_0^ 2
 AL}{8\pi\omega}(1-v/c)\approx\hbar N(1-v/c),
\end{equation}
which is manifestly not Lorentz invariant.

Figure 1 gives a geometric description of this result and
illustrates in a direct way why different photon numbers seem to
appear in different frames.

\begin{figure}[htb]  \centering
\includegraphics[height=4.cm]{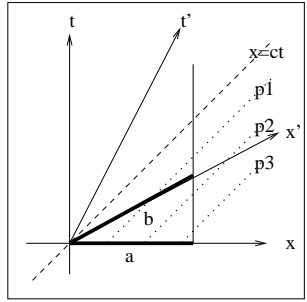}
\caption{Space time diagram. Each photon is
represented by a dotted line (denoted by p1-p3). The solid lines
'a' and 'b' represent the box as viewed at $t=0$, $t'=0$ from the
two frames S and S'. The number of intersections between the
photon world lines and the box gives the total photons inside the
box. It is seen that p2 and p3 are not counted in S', and
therefore there will be more photons counted is S.}
\end{figure}

\section{What Went Wrong?}

What, if anything, went wrong? One easy way out  to say that
photons can only be correctly discussed in a
quantum context. To correctly compute the number of photons one
has  to construct  quantum fields, creation and annihilation operators,
and compute the number of photons
in the framework of quantum field
theory.  It is, of course, correct that a  deeper understanding of
photons requires quantum fields. However, it seems unlikely that this is the only
resolution of a simple paradox. In any case, this is hardly a satisfactory
resolution of it.

The origin of the paradox is not computational or quantum mechanical
 but conceptual.
It all has to do with what is the correct energy $U$ to put in (\ref{N}).
Let us analyze this in some  detail.

Equation~(\ref{N}) must be viewed as a formula that gives the number of
photons in a field configuration at a given instant. A field
configuration is, of course,  extended in space. The field
configuration associated with a plane wave is problematic because
the total electromagnetic energy is infinite, and so is the total
number of photons.  The  energy in a box is finite, however. Yet
the box we  picked is  a virtual box: A box that lets  light
escape and enter. So what we learn is that one can not take a part
of a field configuration  and chop it more or less arbitrarily and
still hope that equation~(\ref{N}) will correctly count the number of
photons. The equation comes with the proviso that the energy is
the total electromagnetic energy of a field configuration. To make
a field configuration with finite energy (and well defined frequency)
 one can confine the electromagnetic field to an
ideal, but still real box. This means a box with reflecting (that
is the real part) and lossless (that is the ideal part) walls. The
field configuration we have picked does not possess this properties.

A second way to resolve the paradox is to think about
equation~(\ref{N}) differently, namely, to think of $U$ as the
energy absorbed by a photo-detector. In this case, the energy $U$
is associated with the energy flux swept by a photo-detector while
it is operating, see figure 2. The relevant box is now not a box
in space but a box in time. The advantage of a detector is that
one can also apply equation~(\ref{N}) to field configurations,
like plane waves, with infinite energy. Since simultaneity  is not
a Lorentz invariant concept,
 extended objects are a pain in special
relativity and a source of many paradoxes. Therefore, a good
photodetector must be a small, and ideally, point-like object.
\begin{figure}[htb]\centering
\includegraphics[height=2.cm]{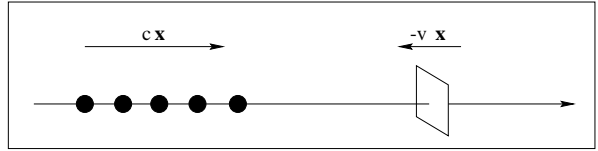}
\caption{A second way to resolve the paradox: The square plate
represents the photodetector, and the dots represent photons.}
\end{figure}

\section{Photons in a Box}
 Photons confined to a box correspond classically to a standing wave.
 A standing wave is
a superposition of two monochromatic waves of equal
frequency and amplitude, traveling in opposite directions.

Let $N_\rightarrow$ and $N_\leftarrow$ denote the number of right-
and left-traveling waves, respectively.
 In the box's rest frame, these
numbers are equal, and we will denote them by N/2. In the moving
frame, the numbers transform according to (\ref{eq:new_n1}):
\begin{equation}
\label{eq:N_R_l}
 N'_\rightarrow=\frac N 2 (1-v/c),\quad
N'_\leftarrow=\frac N 2 (1+v/c)
\end{equation}
Happily, we find $N=N'$ and it is therefore  invariant. So,
although the number of right and left movers are not Lorentz
invariant, their sum is. This is good news, because there are no
additional quantum numbers in this problem besides the total
number of photons.

Although this calculation gives the desired result, it is
cheating: generally, electromagnetic energies do not add linearly.
However, in this case the total energy can be decomposed into two
contributions due to the left- and right-traveling waves. Let $\bf
E_\rightarrow = \hat {\bf y}\rm E_\rightarrow(x,t)$ and $\bf
B_\rightarrow\rm = \hat {\bf z} E_\rightarrow(x,t)$ denote the
electric and magnetic fields of the right-going wave,
respectively. Analogously, the fields of the left-going wave are
$\bf E_\leftarrow\rm = \hat {\bf y}\rm E_\leftarrow(x,t)$ and $\bf
B_\leftarrow\rm = -\hat {\bf z}\rm E_\leftarrow(x,t)$. The sign of
$\bf B_\leftarrow\rm$ is negative because the direction of motion
is reversed. The  energy density is:
\begin{eqnarray}
\label{eq:energy3} &\phantom{=}&\frac{1}{8\pi}\,\Big(\bf E_\rightarrow +
E_\leftarrow\Big)\rm^2 +\frac{1}{8\pi} \Big(\bf B_\rightarrow +
B_\leftarrow\Big)^2\nonumber \\
&=& \frac{1}{8\pi} \,\Big(E_\rightarrow(x,t)
+ E_\leftarrow(x,t)\Big)^2 + \frac{1}{8\pi}\,\Big(E_\rightarrow(x,t) -
E_\leftarrow(x,t)\Big)^2 \nonumber \\ &=& \frac{2}{8\pi}\,\left(E_\rightarrow^2(x,t) +
E_\leftarrow^2(x,t)\right)
\end{eqnarray}
We see that the cross terms  cancel, and the
energies of the two waves indeed add linearly. Note that this
result is true regardless of the reference frame, since we did not
assume any relation between $E_\rightarrow(x,t)$ and
$E_\leftarrow(x,t)$.

Another way of solving the issue of additivity is shown  in figure
3.

\begin{figure}[htb] \centering
\includegraphics[height=1.6cm]{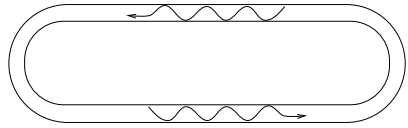}
\caption{Photons in a closed optical fiber. Here, unlike in the
box, photons going in opposite directions don't interfere, and the
energies of the right and left movers are clearly add.}
\end{figure}

\section{Photo-Detector}
A different approach to counting photons in a Lorentz invariant way
is to replace the box by a photodetector. Consider
 a monochromatic plane wave passing through a thin photon
detector whose surface is perpendicular to the x axis, as can be
seen in figure 2.
 We will find the number of photons
passing through the detector during a given proper time $\tau$,
assuming that the photons are point particle.

In the rest frame of the detector, the total energy received by the detector
 during the time $\tau$
is
\begin{equation}
\label{eq:energy2}
 U = \frac{E_0^ 2}{8\pi}Ac\tau
\end{equation}
Where A is the surface area of the detector. This yields
\begin{equation}
\label{eq:n2}
\hbar N = \frac{E_0^2 Ac\tau}{8\pi\omega}
\end{equation}
for the number of photons detected.

In a moving frame the field intensity and frequency
transform according to (\ref{eq:new_e}) and (\ref{eq:new_omega}) respectively.
The measurement time experiences time
dilation:
\begin{equation}
\label{eq:new_time}
t' = \frac{\tau}{\sqrt{1-(v/c)^2}}
\end{equation}
What volume will the detector sweep during $t'$? The detector
moves towards the photons a distance of $vt'$, while each photon,
treated as a point particle, travels towards the detector a
distance of $ct'$. Therefore, the last photon to meet the detector
at time $t'$ is exactly a distance $vt'+ct'$ from the detector at
$t=0$. The volume swept by the detector is $A(v+c)t'$. Now we can
find $N'$:
\begin{eqnarray}
\label{eq:new_n2}
\hbar  N' &=&\frac{(E_0')^ 2}{8\pi}\ A (c+v)\, t'\frac{1}{\omega'}
=\nonumber \\ &=&\frac{E_0^ 2}{8\pi}\,\frac{1-v/c}{1+v/c}\ A (c+v)\,
\frac{\tau}{\sqrt{1-(v/c)^2}} \,
\frac{1}{\omega}\sqrt{\frac{1+v/c}{1-v/c}}\nonumber \\ &=&
A\frac{E_0^ 2 c\tau}{8\pi\omega}=\hbar N
\end{eqnarray}
The number of photons seen in the two frames is the same.
\section{Classical invariants and Ehrenfest Principle}
The point of view which we have taken here, namely that of
examining the significance of  classical  quantities associated
with (discrete) quantum numbers, goes back to Ehrenfest  and the
early days of quantum mechanics  \cite{ehrenfest}. Ehrenfest
stressed the relation of quantum numbers with classical adiabatic
invariants.

Let us recall how this applies to the classical harmonic oscillator.
The ratio of energy to frequency of an oscillator is a  classical
quantity  whose importance in quantum mechanics comes from the fact
that it is a function of the quantum number:
\begin{equation}
\hbar \left(n+\frac{1}{2}\right)=\frac{U}{\omega}
\end{equation}
The ratio of energy to frequency is the classical adiabatic
invariant for the Harmonic oscillator \cite{arnold}. The Ehrenfest
adiabatic principle can be also applied to the quantization of
angular momentum, and the quantization of energy levels in the
Hydrogen atom. Ehrenfest was very specific in identifying
adiabatic invariants with quantum numbers.

It is not clear how to apply the Ehrenfest adiabatic principle to
the number of photons.  However, the Lorentz invariance of the
number of photons
 suggests that one may take a broader interpretation of the Ehrenfest principle
where quantum numbers are associated with a  class of
classical invariants, which includes adiabatic
and Lorentz invariants as special cases.

\section{ Zeldovich Formula}

The question ``Is the number of Photons  a Classical Invariant?"
has been asked, and answered, by Zeldovich in 1966. He pointed out
that the number of photons is both an adiabatic  and a Lorentz
invariant. However, although he made the claim of Lorentz
invariance, he did not {\em show} this.

Zeldovich  wrote an interesting expression for the classical
invariant, which is a generalization of equation~(\ref{N}) to the case
where the field is not monochromatic. The main purpose of this
section is, however, to show that Zeldovich formula indeed
describes a Lorentz invariant, at least for plane waves that are
not monochromatic.

 Zeldovich's
formula for the number of photons is:
\begin{equation}
\label{eq:zel}\hbar  N =\frac{1}{8\pi}\, \int d^3 k\,
\frac{|\hb{E}(\B{k},t)|^2+|\hb{B}(\B{k},t)|^2}{c|\B{k}|}
\end{equation}
which is a natural generalization of equation (\ref{N}) to the
polychromatic case. Here
$\hb{E}(\B{k},t)$ and $\hb{B}(\B{k},t)$ are the Fourier
transforms of the electric and magnetic fields:
\begin{eqnarray}
\label{eq:ekbk} \hb{
E}(\B{k},t)&=&\frac{1}{(2\pi)^{3/2}}\int\,\B{E}(\B{x},t)
e^{-i\B{k} \cdot \B{ x}}\, d^3 x, \nonumber \\
 \hb{B}(\B{k},t)&=&\frac{1}{(2\pi)^{3/2}}\int
\B{B}(\B{x},t) e^{-i\B{k} \cdot \B{ x}}\,d^3x
\end{eqnarray}
 It would be interesting to
have an elementary demonstration of the Lorentz invariance of
equation~(\ref{eq:zel}). Here, instead, we shall be content with the
Lorentz invariance in the special case of a plane wave. That is, a
field configuration that is independent of the $y$ and $z$
coordinate.
 For a plane wave the number of photons
is, of course, infinite, and a finite interesting quantity is the
number of photons per unit area in the $y-z$ plane. Since this
area does not contract under Lorentz boosts in the $\hb{x}$
directions, the corresponding invariant is
\begin{equation}
\hbar n =\frac{1}{8\pi} \int d k\,
\frac{|\hb{E}(k,t)|^2+|\hb{B}(k,t)|^2}{c|k|}
\end{equation}
In the Lorentz frame S a plane wave is made of a right and left
movers:
\begin{eqnarray}
\label{eq:fields}
\B{E}(x,t)&=&(E_\rightarrow(x-ct)+E_\leftarrow(x+ct))\,\hat{\bf
y},\nonumber \\ \B{B}(x,t)&=&
(E_\rightarrow(x-ct)-E_\leftarrow(x+ct))\,\hat{\bf z}
\end{eqnarray}
Using the standard transformation law for the fields
\cite{landau}, one finds that the fields in the frame $S'$, as
functions of its  (primed) coordinates,
\begin{equation}
\label{eq:cortrans} x=\frac{x'+(v/c)t'}{\sqrt{1-(v/c)^2}} \quad
t=\frac{t'+(v/c)x'}{\sqrt{1-(v/c)^2}}
\end{equation}
to be
\begin{eqnarray}
\label{eq:newfields} \B{E'}(x',t')&=&\left(b\,E_\rightarrow
(b(x'-ct'))+ \frac{1}{b}\,E_\leftarrow\left( \frac{x'+c
t'}{b}\right)\right)\,\hb{y'}\nonumber
\\
\B{B'}(x',t')&=&\left(b\,E_\rightarrow (b(x'-ct'))
-\frac{1}{b}\,E_\leftarrow\left(\frac{x'+c
t'}{b}\right)\right)\,\hb{z'}
\end{eqnarray}
where $b\equiv\sqrt{\frac{1-v/c}{1+v/c}}$. We can now compute
 $\hb{E}'( k',t')$ and $\hb{B}'( k',t')$.
\begin{eqnarray}
\label{eq:newek} \hb{E}'( k',t' )&=&
\hb{y'}\frac{1}{(2\pi)^{3/2}}\int\,\left( b E_\rightarrow
(bx')+\frac{1}{b}\, E_\leftarrow\left(\frac{x'}{b}\right)\right)
\,e^{-i k' x'}dx'\nonumber
\\ &=&\hb{y'}\left(\hat
{E}_\rightarrow\left(\frac{k'}{b}\right)+\hat {E}_\leftarrow(k'
b)\right)
\end{eqnarray}
and similarly
\begin{equation}
\label{eq:newbk} \hb{B}'(k',t' )= \hb{z}'\,\left({\hat
E_\rightarrow}\left(\frac{k'}{b}\right)- {\hat E_\leftarrow}(k'
b)\right)
\end{equation}
The number of photons per unit area in the  frame $S'$ is
\begin{eqnarray}
\hbar n'&=&\frac{2}{8\pi}\, \int{\left(\left|\hat
E_\rightarrow\left(\frac{k'}{b}\right)\right|^2+ \left|\hat
E_\rightarrow(k'b)\right|^2\right) \frac{dk'}{c|k'|}}\nonumber
\\ &=&\frac{2}{8\pi}\,\int{\left(|\hat E_\rightarrow(k)|^2 +
|\hat E_\rightarrow(k)|^2\right)\frac{dk}{c|k|}}=\hbar n
\end{eqnarray}
where the factor $2$ comes from the contribution of the electric
and magnetic fields, and the mixed terms drop. This
establishes  Lorentz invariance.

\section{Epilogue}
This is an account of a simple paradox and its resolution. It is
remarkable that in spite of the quantum nature of photons, one can
correctly compute their number in a box as if they were mere golf
balls is a bucket.  However, precisely because photons are, at the
same time, associated with an extended field configuration, this
calculation is also subtle, and can lead to wrong results if one
is not careful.

The account given here grew out  of teacher-students interaction in the spring
semester class of classical electrodynamics at the Technion.  Puzzles
 are effective means to teach and learn especially
when the teacher does not already know the resolution.

\ack We thank Prof.~Andrew Zangwill for drawing our attention to
the work of Zeldovich. This work is supported in part by the
Israel Academy of Sciences, the DFG, and by the Fund for the
Promotion of Research at the Technion.

{\bf Note Added:} After this paper has been sent to the printer,
we were kindly informed by  Jean-Jacques Labarthe, that the
problem was first considered by by A. Einstein in {\it Zur
Elektrodynamik bewegter K\"orper}, Ann. Phys. {\bf 17}, 891-921,
(1905). See also A. I. Miller, {\it Albert Einstein Special Theory
of Relativity}, Addison Wesley, (1981).

\end{document}